---

**Contribution aux travaux du Comité interministériel de l'intelligence artificielle générative**

« Recommandations pour une action publique en faveur d'une IA générative respectueuse de l'environnement »

Thomas Le Goff

*Docteur en droit, Université Paris Cité (CEDAG – EA1516)*
*Electricité de France (EDF SA), Direction Juridique*

---

*Les opinions exprimées dans la présente contribution n'engagent que leur auteur et non les organisations auxquelles il est rattaché.*

Résumé :

La prise de conscience croissante de l'impact environnemental des technologies numériques a donné lieu à plusieurs initiatives isolées visant à promouvoir des pratiques durables. Cependant, malgré ces efforts, l'empreinte environnementale de l'IA générative, notamment en termes d'émissions de gaz à effet de serre et de consommation d'eau, reste considérable. La présente contribution aborde d'abord les composantes de cette empreinte environnementale, mettant en évidence les émissions massives de CO2 et la consommation d'eau associées à l'entraînement des grands modèles de langage, soulignant ainsi la nécessité de repenser les méthodes d'apprentissage et d'inférence pour réduire cet impact. La contribution explore également les facteurs et caractéristiques des modèles ayant une influence sur leur empreinte environnementale et démontre l'existence de solutions pour minimiser cette dernière, telles que l'utilisation de processeurs plus performants ou l'optimisation de la performance énergétique des centres de données. Les potentiels effets néfastes de l'IA sur la planète et son écosystème ont conduit la protection de l'environnement au rang des principes fondateurs de l'éthique de l'IA au niveau international ou européen. Toutefois, cette reconnaissance du problème ne s'est pour le moment pas traduite par la prise de mesures concrètes pour y répondre.

A ce titre, la présente contribution propose douze recommandations pragmatiques pour une action publique visant à promouvoir une IA générative durable, notamment en construisant une stratégie à long terme pour atteindre la neutralité carbone des modèles d'IA, en encourageant la coopération internationale pour l'édiction de normes communes, en soutenant la recherche scientifique et en élaborant des cadres juridiques et réglementaires adaptés.

La présente contribution cherche ainsi à éclairer les membres du Comité interministériel de l'IA générative sur les enjeux environnementaux de cette technologie en leur proposant un bref état de l'art de la littérature scientifique en la matière et en proposant des pistes d'action concrètes pour concilier l'innovation technologique avec la nécessaire préservation de l'environnement.

**Introduction :**

L'empreinte des technologies numériques sur l'environnement n'est pas une préoccupation nouvelle et a déjà fait l'objet d'une littérature abondante (ADEME, 2021 ; CNNum, 2021 ; Dhar, 2020). Cette prise de conscience s'est jusqu'ici traduite par la création de collectifs visant à promouvoir le développement de technologies numériques durables[1] ou par de premières initiatives législatives comme la loi « Réduire l'Empreinte Environnementale du Numérique » du 15 novembre 2021 (Fonbaustier, 2022).

L'empreinte environnementale du numérique est généralement décomposée entre les émissions de gaz à effet de serre générées par l'extraction des matériaux nécessaires à la production des terminaux physiques, l'alimentation énergétique des centres de données ou encore la consommation en eau de ces derniers (Sénat, 2020). C'est dans ce contexte d'empreinte environnementale encore mal maîtrisée que l'IA générative connait un essor sans précédent. En effet, les besoins en capacité de calcul (et donc en énergie) des *large language models* (LLM) sont colossaux, de l'ordre de 1 287 000 kWh pour le pré-entraînement de GPT-3 (Patterson *et al.*, 2021).

Le développement exponentiel des LLM, la multiplication des modèles et l'absence de maîtrise de l'empreinte environnementale ont conduit de nombreux chercheurs à tirer la sonnette d'alarme et à recommander le développement de méthodes plus respectueuses de l'environnement (McDonald *et al.*, 2022).

Si les initiatives de régulation portant sur l'éthique du développement de l'IA se multiplient, la question de son empreinte environnementale est bien souvent reléguée au second plan. Les différentes chartes éthiques ou prises de position politiques témoignent pourtant d'une prise de conscience sur cet enjeu.

La création du comité interministériel de l'IA générative par le gouvernement français est une initiative bienvenue pour contribuer au développement éthique de cette technologie en France et en Europe. La présente contribution entend apporter aux membres du comité un éclairage sur l'épineuse question de l'empreinte environnementale de l'IA générative. Loin de vouloir restreindre l'innovation, nous cherchons plutôt à identifier des actions concrètes pour encourager le marché à adopter des pratiques responsables dans une logique d'autorégulation.

Pour parvenir à cet objectif, la contribution donnera d'abord au lecteur un bref état de l'art et des chiffres clés sur l'empreinte environnementale de l'IA générative (**I**). Ensuite, il sera démontré que la prise en compte de la protection de l'environnement dans le développement de l'IA générative apparaît nécessaire au regard des principaux textes de *soft law* sur l'éthique de l'IA à l'échelle internationale (**II**). Enfin, des recommandations concrètes seront proposées, lesquelles pourraient être librement reprises par le Comité dans ses travaux (**III**). Ces propositions visent à engager la France dans une posture de *leader* en matière d'éthique environnementale de l'IA, en promouvant le développement de technologies limitant leur impact sur la planète tout en favorisant les cas d'usage contribuant aux efforts de lutte contre le changement climatique.

---

[1] Voir notamment le collectif « Green IT » (https://www.greenit.fr/) et ses applications dans les entreprises (CIGREF, 2017).

## I – L'empreinte environnementale de l'IA générative : état de l'art

La recherche sur l'empreinte environnementale de l'IA s'est développée au cours de la dernière décennie, accompagnant la multiplication des cas d'usage. Cette empreinte comprend trois composantes principales (**A**) et son importance est définie par la taille et les caractéristiques du modèle considéré (**B**). Toutefois, la recherche a également mis en évidence l'existence de mesures techniques dont la mise en œuvre permettrait de réduire les besoins énergétiques des modèles d'IA génératives (**C**).

### A/ Les composantes de l'empreinte environnementale de l'IA générative

Tout d'abord, l'IA présente une empreinte « carbone » importante. Elle est constituée par les émissions de gaz à effet de serre liées en grande partie à la consommation énergétique requise pour les opérations de calcul (entraînement des modèles d'apprentissage automatique, inférences…). La source d'énergie utilisée a évidemment un impact sur l'intensité carbone des calculs donc il est difficile de mesurer précisément cette empreinte, que l'on sait pourtant très importante. La question de la consommation en électricité des centres de données a fait l'objet de nombreuses recherches au cours de la dernière décennie (Jones, 2018). Pour ce qui concerne l'IA générative, une étude du MIT a révélé en 2019 que la seule phase d'entraînement d'un LLM de 1,75 milliards de paramètres pouvait nécessiter plus de 650 000 kWh d'électricité et conduire à l'émission de 280 tonnes de CO2 (Strubell *et al.*, 2019). Pour comparaison, la consommation moyenne d'un foyer de deux personnes en France est d'environ 4679 kWh par an[2]. L'entraînement du modèle de 1,75 milliards de paramètres évoqué ci-dessus représenterait la consommation annuelle d'électricité d'environ 140 foyers et il faut garder à l'esprit que les nouveaux modèles de langue présentent un nombre bien plus important de paramètres, comme GPT-3 avec 175 milliards de paramètres.

| Model | Hardware | Power (W) | Hours | kWh·PUE | $CO_2e$ | Cloud compute cost |
|---|---|---|---|---|---|---|
| Transformer$_{base}$ | P100x8 | 1415.78 | 12 | 27 | 26 | $41–$140 |
| Transformer$_{big}$ | P100x8 | 1515.43 | 84 | 201 | 192 | $289–$981 |
| ELMo | P100x3 | 517.66 | 336 | 275 | 262 | $433–$1472 |
| BERT$_{base}$ | V100x64 | 12,041.51 | 79 | 1507 | 1438 | $3751–$12,571 |
| BERT$_{base}$ | TPUv2x16 | — | 96 | — | — | $2074–$6912 |
| NAS | P100x8 | 1515.43 | 274,120 | 656,347 | 626,155 | $942,973–$3,201,722 |
| NAS | TPUv2x1 | — | 32,623 | — | — | $44,055–$146,848 |
| GPT-2 | TPUv3x32 | — | 168 | — | — | $12,902–$43,008 |

Table 3: Estimated cost of training a model in terms of $CO_2$ emissions (lbs) and cloud compute cost (USD).[7] Power and carbon footprint are omitted for TPUs due to lack of public information on power draw for this hardware.

E. Strubell, A. Ganesh, A. Mccallum, "Energy and Policy Considerations for Deep Learning in NLP", *57th Annual meeting of the Association for Computational Linguistics*, 5 juin 2019, https://arxiv.org/pdf/1906.02243.pdf.

---

[2] ElecDom – Données de consommation annuelle, jeu de données de l'ADEME publié sur DataGouv : https://www.data.gouv.fr/fr/datasets/elecdom-donnees-de-consommation-annuelle/.

Ensuite, l'IA dispose d'une empreinte environnementale liée à sa consommation en eau. Des chercheurs de l'Université de Californie ont démontré en 2023 que l'entraînement de GPT-3 dans les centres de données de Microsoft aux États-Unis a conduit à la consommation de près de 700 000 litres d'eau douce et potable. Leur étude montre également que l'IA pourrait être responsable de 4,2 à 6,6 milliards de mètres cubes de prélèvement d'eau en 2027, soit une consommation annuelle d'eau équivalente à quatre ou six fois celle du Danemark, ou la moitié du Royaume-Uni (Li *et al.*, 2023 ; Ren, 2020). Le tableau ci-dessous représente la consommation en eau estimée pour l'entraînement de GPT-3 et par inférence, fondée sur les données affichées par Microsoft sur la consommation de ses centres de données.

*Table 1: Estimate of GPT-3's average operational water consumption footprint. "*" denotes data centers under construction as of July 2023, and the PUE and WUE values for these data centers are based on Microsoft's projection.*

| Location | PUE | WUE (L/kWh) | Electricity Water Intensity (L/kWh) | Water for Training (million L) | | | Water for Each Inference (mL.) | | | # of Inferences for 500ml Water |
|---|---|---|---|---|---|---|---|---|---|---|
| | | | | On-site Water | Off-site Water | Total Water | On-site Water | Off-site Water | Total Water | |
| U.S. Average | 1.170 | 0.550 | 3.142 | 0.708 | 4.731 | **5.439** | 2.200 | 14.704 | **16.904** | 29.6 |
| Wyoming | 1.125 | 0.230 | 2.574 | 0.296 | 3.727 | **4.023** | 0.920 | 11.583 | **12.503** | 40.0 |
| Iowa | 1.160 | 0.190 | 3.104 | 0.245 | 4.634 | **4.879** | 0.760 | 14.403 | **15.163** | 33.0 |
| Arizona | 1.223 | 2.240 | 4.959 | 2.883 | 7.805 | **10.688** | 8.960 | 24.259 | **33.219** | 15.1 |
| Washington | 1.156 | 1.090 | 9.501 | 1.403 | 14.136 | **15.539** | 4.360 | 43.934 | **48.294** | 10.4 |
| Virginia | 1.144 | 0.170 | 2.385 | 0.219 | 3.511 | **3.730** | 0.680 | 10.913 | **11.593** | 43.1 |
| Texas | 1.307 | 1.820 | 1.287 | 2.342 | 2.165 | **4.507** | 7.280 | 6.729 | **14.009** | 35.7 |
| Singapore | 1.358 | 2.060 | 1.199 | 2.651 | 2.096 | **4.747** | 8.240 | 6.513 | **14.753** | 33.9 |
| Ireland | 1.197 | 0.030 | 1.476 | 0.039 | 2.274 | **2.313** | 0.120 | 7.069 | **7.189** | 69.6 |
| Netherlands | 1.158 | 0.080 | 3.445 | 0.103 | 5.134 | **5.237** | 0.320 | 15.956 | **16.276** | 30.7 |
| Sweden | 1.172 | 0.160 | 6.019 | 0.206 | 9.079 | **9.284** | 0.640 | 28.216 | **28.856** | 17.3 |

P. Li, J. Yang, M.A. Islam, S. Ren, "Making AI Less "Thirsty": Uncovering and Addressing the Secret Water Footprint of AI Models", *arxiv*, October 2023, https://arxiv.org/abs/2304.03271.

Enfin, la multiplication des usages de l'IA et l'augmentation de la taille des modèles utilisés conduisent à l'accroissement des besoins en infrastructures et, par conséquent, en matériaux. En effet, la conception des ordinateurs sur lesquels fonctionnent les modèles d'IA (principalement en centres de données, largement concentrés entre les mains de quelques acteurs seulement) nécessite des matériaux dits « terres rares » (sillicium, cobalt…), dont l'extraction et le raffinage entraînent des coûts environnementaux et humains considérables (Crawford and Joler, 2018). La problématique n'est pas spécifique à l'IA générative et concerne tout le secteur du numérique (Hilty *et al.*, 2011 ; Plepys, 2002). Toutefois, les importants besoins des LLM en capacités de calcul risquent d'accélérer le phénomène et appellent donc à une prise en compte *ex-ante*.

Au regard des données disponibles sur les besoins énergétiques, en eau et en matériaux rares liés au développement de l'IA, et d'autant plus de l'IA générative reposant sur des modèles de plus en plus imposants, il semble que l'adoption actuelle de cette technologie se fasse en dépit de toute considération écologique. Il est néanmoins important d'identifier les facteurs influant sur cette empreinte. Parmi eux, on retrouve naturellement la performance énergétique des centres de données ou serveurs sur lesquels fonctionnent les modèles ainsi que l'origine de l'énergie les alimentant (électricité provenant d'énergies fossiles ou de sources bas carbone telles que le nucléaire ou les énergies renouvelables). En plus de ces considérations systémiques, des facteurs techniques propres à chaque modèle et à la méthode employée pour les entraîner entrent en jeu.

### B/ L'influence de la taille et du type de modèle d'IA générative sur l'empreinte environnementale

Il serait faux de considérer par défaut que tous les modèles d'IA générative présentent une empreinte environnementale déraisonnable. On constate en effet de fortes disparités entre les modèles, si bien que la littérature scientifique a vu l'apparition d'une distinction entre le « Green AI », systèmes minimisant leur empreinte environnementale, et le « Red AI », systèmes dont l'impact sur la planète est important, que ce soit en raison des émissions de gaz à effet de serre liées à leur apprentissage et à leur fonctionnement, à leur consommation en eau ou à la finalité pour laquelle ils sont utilisés (Dhar, 2020).

La mesure de l'empreinte environnementale de l'IA générative n'est pas aisée tant il y a de facteurs à prendre en compte. C'est la raison pour laquelle les principales recherches publiées se concentrent sur l'évaluation de l'empreinte carbone liée à la consommation énergétique nécessaire aux phases d'apprentissage et d'inférence. Une étude réalisée par des chercheurs de Google et de l'Université de Berkeley a estimé, en 2021, les émissions de CO2 générées par l'entraînement de cinq LLM de tailles différentes (Patterson *et al.*, 2021), dont GPT-3. Ses résultats montrent une forte variabilité du niveau d'émissions entre les modèles, allant de 4 tonnes de CO2 à plus de 550 tonnes pour GPT-3 (V100).

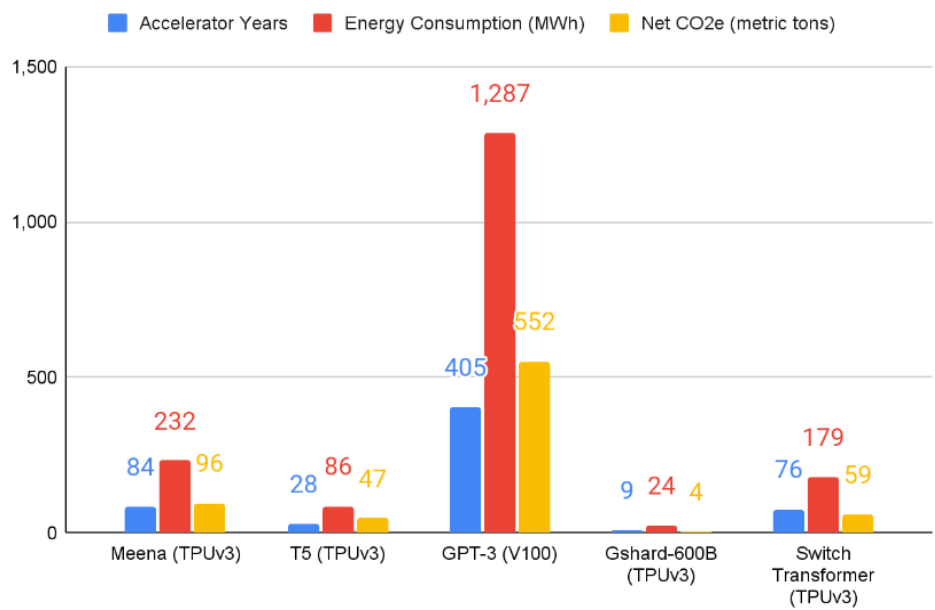

**Figure 3. Accelerator years of computation, energy consumption, and CO$_2$e for five large NLP DNNs.**

D. Patterson, J. Gonzalez, Q. Le, C. Liang, L.-M. Munguia, D. Rothchild, D. So, M. Texier, J. Dean, "Carbon Emissions and Large Neural Network Training", *arxiv*, 2021, https://arxiv.org/pdf/2104.10350.pdf.

Les émissions de gaz à effet de serre étant liées à la consommation électrique requise pour l'entraînement ou le fonctionnement du modèle, elles sont proportionnelles aux besoins en capacité de calcul des modèles et donc à leur nombre de paramètres. Le tableau ci-dessous illustre ce lien en présentant la puissance de calcul requise pour différents types de modèles présentant un nombre de paramètres croissant.

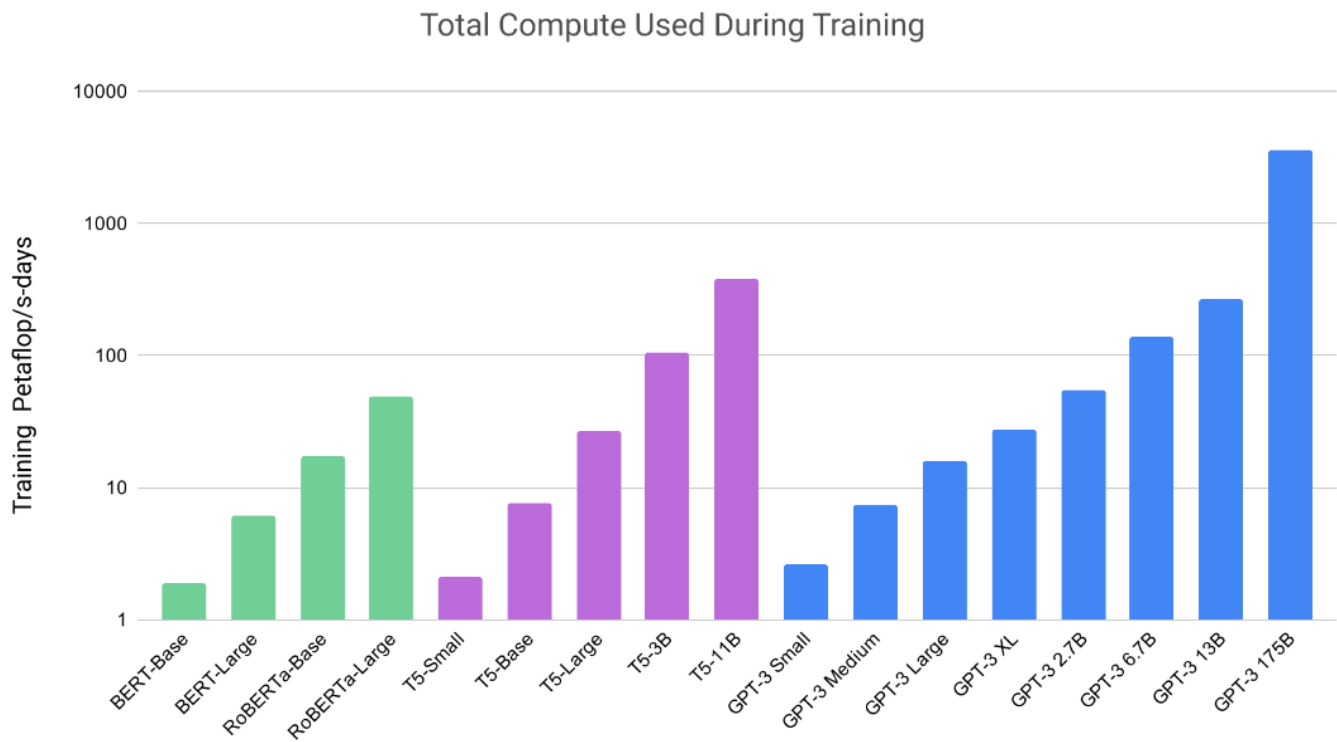

T. Brown, B. Mann, N. Ryder, et al., "Language Models are Few-Shot Learners", *NIPS'20: Proceedings of the 34th International Conference on Neural Information Processing Systems*, December 2020, n°159, p. 1877–1901, https://arxiv.org/pdf/2005.14165.pdf.

Notons à ce stade que, jusqu'à aujourd'hui, la performance des modèles a été corrélée avec leur taille. GPT-2 d'OpenAI utilisait 1,5 milliards de paramètres, tandis que GPT-3 en utilise jusqu'à 175 milliards. Il n'est pas étonnant qu'OpenAI et Google n'aient pas divulgué le nombre de paramètres de leurs derniers modèles les plus performants, respectivement GPT-4 et Gemini Ultra. Les dix dernières années ont été marquées par une augmentation constante des besoins en capacité de calcul pour l'IA (open AI, 2018) et, pour le moment, rien ne laisse supposer une inversion de tendance pour le futur.

Toutefois, malgré ces constats alarmants, il convient de souligner qu'il existe de nombreuses solutions pour minimiser l'empreinte environnementale de l'IA générative, allant de l'optimisation de la performance énergétique des infrastructures de calcul à l'utilisation de modèles plus efficients.

### C/ La faisabilité technique de la minimisation de l'empreinte environnementale de l'IA générative

La suite de nos développements se concentrera sur la réduction des émissions de gaz à effet de serre liées à l'apprentissage et au fonctionnement de l'IA générative, obtenue via la réduction de la capacité de calcul mobilisée. Seront présentés quatre leviers sur lesquels il est possible d'agir : la conception des algorithmes (volet logiciel), les processeurs utilisés, la performance énergétique des centres de données et l'origine de l'électricité consommée (Patterson et al., 2021). L'objectif de cette section n'est pas de présenter une liste exhaustive des mesures qu'il est possible de mettre en place pour assurer l'écoresponsabilité « *by design* » de l'IA générative mais simplement de démontrer que de telles mesures existent.

Premièrement, l'optimisation des algorithmes et des programmes sous-jacents à l'IA générative permettrait de réduire la capacité de calcul nécessaire à l'entraînement des modèles, ce qui conduirait à des économies (i) de temps, (ii) financières et (iii) en termes d'émissions de gaz à effet de serre. Cette optimisation peut passer par plusieurs moyens. D'une part, puisque la consommation énergétique requise est fonction de la taille du modèle, il conviendrait de minimiser la taille du modèle utilisé au regard de la finalité poursuivie. En effet, il n'est pas nécessaire d'utiliser un modèle de plusieurs dizaines de milliards de paramètres pour certaines tâches simples ou pour lesquelles un degré de précision moindre est acceptable : certains modèles plus petits peuvent présenter des performances suffisantes[3]. D'autre part, il est possible d'employer des méthodes dites d'*Early stopping on inference* pour optimiser les besoins en énergie d'un modèle en phase de fonctionnement. Ces méthodes de conception visent à éviter que le modèle soit contraint, pour chaque inférence, de passer par toutes les couches d'un réseau de neurones avant de prendre une décision et de la fournir en résultat. Le principe est d'évaluer le degré de confiance dans le résultat à chaque couche et d'arrêter le processus dès qu'un niveau suffisant de confiance est atteint. L'arrêt prématuré du processus d'inférence permettrait ainsi d'économiser du temps et les coûts associés, tout en garantissant un niveau de performance satisfaisant (voir notamment Zhou *et al.*, 2020). Il ne s'agit là que d'exemples, démontrant l'existence de possibilités pour réduire l'empreinte environnementale de l'IA générative par des choix de conception.

Deuxièmement, la consommation énergétique de l'IA générative peut également être diminuée en ayant recours à du matériel informatique, des processeurs, plus performants. Il ressort de la littérature qu'il existe des types de processeurs plus ou moins adaptés aux types de calculs nécessaires à l'entraînement et le fonctionnement des réseaux de neurones profonds. A ce titre, une partie de l'industrie est passée de l'utilisation de puces GPU (*Graphics Processing Unit*) aux puces TPU (*Tensor Processing Unit*), plus performantes pour l'apprentissage automatique[4]. Toutefois, les types de puces sont plus ou moins adaptées en fonction de la tâche à réaliser. Il convient donc systématiquement de choisir, pour chaque modèle, les ressources informatiques les plus adaptées et performantes pour minimiser la consommation y relative (Jouppi *et al.*, 2021).

Troisièmement, au-delà du matériel informatique en lui-même, il est possible d'optimiser la performance énergétique du centre de données dans lequel il est hébergé (dans le cas du recours à une infrastructure d'informatique en nuage notamment). Cette performance est mesurée par l'indicateur d'efficacité énergétique (IEE)[5] soit le rapport entre l'énergie consommée globalement par le centre de données et l'énergie effectivement consommée par le matériel informatique. Le respect des normes de référence telles que les normes ISO 14001 (système de management environnemental des organisations) et ISO 50001 (management de l'énergie) pour les centres de données hébergeant les modèles d'IA génératives permettrait de garantir une maîtrise de leur empreinte environnementale.

---

[3] Voir notamment les résultats du dernier modèle de Mistral AI comprenant 7 milliards de paramètres, dont les performances outrepassent les modèles Llama 1 et 2 comprenant respectivement 34 et 13 milliards de paramètres (https://mistral.ai/news/announcing-mistral-7b/).
[4] Voir notamment : Google Cloud, "Présentation de Cloud TPU", https://cloud.google.com/tpu/docs/intro-to-tpu?hl=fr.
[5] Traduction libre de *Power Usage Effectiveness (PUE)*.

Quatrièmement, l'empreinte carbone de l'IA générative pourrait être grandement réduite s'il était possible de garantir qu'elle soit alimentée par un mix énergétique décarboné. Le recours à une électricité produite à partir d'énergies nucléaire, renouvelable ou hydraulique limiterait grandement l'empreinte des centres de données hébergeant les modèles d'IA. Inciter les développeurs de grands modèles d'IA génératives à garantir la neutralité carbone de leur approvisionnement en électricité, soit par l'achat de garanties d'origine soit par la conclusion de Power Purchase Agreement (PPA) pour des capacités équivalentes à celles nécessaires à l'opération des modèles d'IA, pourrait avoir un impact significatif sur l'empreinte environnementale du secteur[6].

Cette liste non exhaustive démontre qu'il existe plusieurs façons de minimiser l'empreinte carbone de l'IA générative. Des recherches supplémentaires seraient nécessaires pour déterminer les méthodes les plus efficaces et développer, en conséquence, des standards auxquels pourraient se référer les acteurs du secteur. Une telle démarche devrait également être entreprise pour l'impact de l'IA générative en termes de consommation en eau et de matériaux.

Au vu de l'ensemble des constats présentés jusqu'ici et des solutions envisageables, la prise en compte de la protection de l'environnement dans la construction d'une éthique de l'IA générative apparait indispensable.

## II – La protection de l'environnement : une composante essentielle de l'éthique de l'IA générative

Partout dans le monde, les réflexions sur l'éthique de l'IA foisonnent. De nombreuses lignes directrices, référentiels, chartes et autres textes de *soft law* ont ainsi été publiés au cours des dix dernières années. Une équipe de chercheurs suisses, à l'issue d'un travail de cartographie réalisé en 2019, a démontré une convergence globale des textes sur l'éthique de l'IA autour de cinq principes : la transparence, la justice et l'équité, la sécurité, la responsabilité et la protection de la vie privée (Jobin *et al.*, 2019). Toutefois, la durabilité environnementale de l'IA et l'idée qu'une IA éthique serait une IA garantissant la protection de l'environnement, de la biodiversité et des écosystèmes, se retrouve également dans de nombreux textes de référence cités par l'étude.

Au niveau international, l'enjeu de la protection de l'environnement dans l'éthique de l'IA se retrouve à la fois dans la Recommandation du Conseil de l'OCDE sur l'IA du 22 mai 2019 (OCDE, 2019), adoptée par 42 pays dont la France, et dans la Recommandation sur l'éthique de l'IA de l'UNESCO adoptée en 2021 (UNESCO, 2021). Cette recommandation contient notamment un principe invitant à mettre les services d'IA au service de la prospérité de l'environnement et des écosystèmes, exprimé très clairement : « *Tous les acteurs impliqués dans le cycle de vie des systèmes d'IA [...] devraient réduire l'impact environnemental des systèmes d'IA, ce qui inclut, sans s'y limiter, leur empreinte carbone, afin de réduire autant que possible les facteurs de risque associés au changement climatique et aux changements*

---

[6] Voir notamment l'engagement de Google sur la neutralité carbone : Google, "24/7 Carbon-Free Energy: Methodologies and Metrics", February 2021, https://www.gstatic.com/gumdrop/sustainability/24x7-carbon-free-energy-methodologies-metrics.pdf.

*environnementaux, et d'empêcher l'exploitation, l'utilisation et la transformation non durables des ressources naturelles, qui contribuent à la détérioration de l'environnement et à la dégradation des écosystèmes* »[7]. À date, aucun instrument normatif contraignant n'a suivi l'adoption de ces principes.

Au niveau européen, ensuite, le livre blanc sur l'IA de la Commission européenne du 19 février 2020 (CE, 2020)[8] ou la résolution du Parlement européen relative aux aspects éthiques de l'IA, la robotique et autres technologies du 20 octobre 2020[9] soulignent également cet enjeu écologique. Pourtant, alors même que le « bien-être environnemental » figure bien au rang des sept principes éthiques en matière d'IA auxquels adhèrent les institutions européennes (HLEG, 2019), le projet de règlement européen sur l'IA[10] ne comporte aucune disposition visant à limiter les conséquences environnementales néfastes du développement de cette technologie.

Enfin, le constat est identique à l'échelle nationale. Le dilemme écologique du développement de l'IA a été évoqué dès les travaux du député Cédric Villani dans son rapport de 2018 (Villani, 2018), mais, depuis, aucune mesure concrète n'a été prise. On notera tout de même la publication en juillet 2020 de la Feuille de route sur l'environnement et le numérique, coconstruite par le Conseil national du numérique et le Haut conseil pour le climat (CNNum, 2020). Y figurent de nombreuses mesures visant la construction d'un numérique sobre, au service de la transition écologique et solidaire. Ces mesures s'articulent autour de trois grands axes, particulièrement pertinents au regard de notre sujet : le premier appelle à réduire l'empreinte environnementale du numérique, le deuxième à mobiliser le potentiel du numérique au service de la transition écologique, et le dernier à accompagner l'ensemble de la société vers un numérique responsable. Il faut également citer les travaux de l'ARCEP sur le numérique et l'environnement (ARCEP, 2023) ainsi que la feuille de route du Ministère de la transition écologique contenant des objectifs spécifiques à l'IA frugale (MTE, 2023). Ne contenant pas de dispositions contraignantes et étant relativement récentes, ces publications n'ont pour le moment donné lieu qu'à peu de mesures concrètes. Pourtant, bon nombre des propositions qu'elles contiennent sont essentielles pour la construction d'un cadre durable au développement de l'IA, objet de notre propos.

Pour ce qui concerne l'IA générative, ce défaut d'actions concrètes peut s'expliquer de plusieurs manières. D'abord, pendant longtemps, l'absence de données précises sur cette empreinte a rendu difficile la prise de conscience du public et des pouvoirs publics (Dhar, 2020). Ensuite, lorsqu'un risque n'est pas immédiat (faute de conséquences en cas d'empreinte trop élevée) et avéré (faute de données fiables et consolidées sur les conséquences d'une empreinte élevée), les responsables politiques et les entreprises du secteur privé tardent à agir. Enfin, le développement de l'IA générative est imprévisible et très rapide, ce qui rend vaine toute tentative visant à figer un cadre de régulation. La faible maturité de la technologie et la multiplication des modèles publiés ces dernières années constituent ainsi de véritables obstacles à la création d'une règlementation contraignante.

---

[7] UNESCO, 2021, point 18.
[8] Notamment en page 4 : « *L'importance de l'IA ne cessant de croître, il faut dûment tenir compte de l'incidence environnementale des systèmes d'IA tout au long de leur cycle de vie et sur l'ensemble de la chaîne d'approvisionnement, c'est-à-dire en ce qui concerne l'utilisation des ressources pour l'entraînement des algorithmes et le stockage des données.* »
[9] *Résolution 2020/2012(INL) du Parlement européen du 20 octobre 2020 portant recommandations à la Commission sur un cadre aux aspects éthiques de l'intelligence artificielle, la robotique et autres technologies*, p. 14, N°51, notamment : « *the development, deployment and use of these technologies should contribute to the green transition, preserve the environment, and minimise and remedy any harm caused to the environment during their lifecycle and across their entire supply chain.* »
[10] COMMISSION EUROPÉENNE, *Proposition de règlement du Parlement européen et du Conseil établissant des règles harmonisées concernant l'intelligence artificielle (législation sur l'intelligence artificielle) et modifiant certains actes législatifs de l'union*, 21 avril 2021, 2021/0106 (COD).

Toutefois, cet argument, dont la véracité ne peut être remise en cause, ne saurait justifier l'inaction des pouvoirs publics. Il existe de nombreux leviers sur lesquels l'action publique pourrait agir afin de garantir une éthique environnementale de l'IA générative. Figurant dans les priorités de la plupart des textes internationaux en matière d'éthique de l'IA, comme cela a été présenté ci-dessus, la protection de l'environnement ne peut en rester le parent pauvre.

## III – Recommandations pour une action publique en faveur d'une IA générative durable en France et en Europe

Développer une éthique de l'IA générative doit impérativement inclure des actions pour assurer un développement technologique compatible avec les objectifs de développement durable, de sobriété énergétique, de neutralité carbone. Il n'est pas ici question de proposer un cadre de régulation exhaustif, ce qui, comme cela a été expliqué, serait difficile compte tenu de l'évolutivité de la technologie et nécessiterait un effort de recherche coordonné et international. La présente contribution propose plutôt des recommandations d'actions concrètes et réalistes pour contribuer à développer la recherche sur l'éthique environnementale de l'IA générative ainsi que pour promouvoir l'adoption volontaire de pratiques vertueuses en la matière.

Ces propositions sont propres à leur auteur ou fondées sur plusieurs travaux de recherche académique (notamment Kaack *et al.*, 2022 ; Stein, 2020 ; Hacker, 2023). Elles s'inscrivent dans la thématique de travail du Comité interministériel de l'IA générative portant sur *« L'éthique et les impacts sociétaux, notamment la façon d'adapter l'encadrement de ces technologies pour qu'elles soient effectivement au service de nos valeurs, sans brider l'innovation ».*

### Volet « Neutralité carbone de l'IA générative »

- **Favoriser la création de jeux de données de référence, validés et publiés en *open data*, sur l'empreinte environnementale de l'IA générative**, en ayant notamment recours aux compétences et expertises de l'ARCEP dans le prolongement de ses travaux sur le numérique et l'environnement.

- **Développer une feuille de route interministérielle fixant des objectifs à court, moyen et long termes pour atteindre la neutralité carbone de l'IA générative** sur le territoire français et européen.

- **Favoriser la prise en compte de l'empreinte environnementale dès la conception (« sustainability by design ») : établir et diffuser des bonnes pratiques** permettant la mesure de l'empreinte des IA génératives et le développement d'applications frugales (fiches pratiques de sensibilisation à l'impact environnemental des choix de conception, de la taille du modèle utilisé, des techniques de développement employées…), sur le modèle du référentiel de l'écoconception des produits numériques de l'ARCEP.

- **Contribuer à la prise de conscience du grand public sur l'empreinte environnementale de l'IA générative en publiant des contenus pédagogiques**.

**Volet « Coopération internationale »**

- Se positionner en *leader* et **porter la dimension environnementale dans les réflexions internationales sur l'éthique de l'IA** et les initiatives de régulation (OCDE, UNESCO, GPAI, Union européenne…).

**Volet « Recherche scientifique »**

- Poursuivre et amplifier la promotion des solutions d'IA générative respectueuses de l'environnement, minimisant leur impact et contribuant à la transition écologique, par le financement de la recherche :
    - o **Soutenir la recherche sur des projets d'IA générative contribuant à la lutte contre le réchauffement climatique et à la réduction des émissions de gaz à effet de serre** (conseil aux particuliers pour les économies d'énergie, analyse automatisée des dossiers d'autorisation pour des projets d'installation de moyens de production renouvelables…)
    - o **Soutenir la recherche sur la minimisation de l'empreinte environnementale de l'IA générative**

- **Soutenir la recherche sur l'évaluation de l'empreinte environnementale de l'IA générative dans le but de développer des standards** français ou européens.

**Volets « Juridique et règlementaire »**

- **Mettre en place un dispositif étatique d'accompagnement volontaire sur le modèle des « bacs à sable règlementaires » permettant à des projets d'IA générative de bénéficier d'un accompagnement renforcé des autorités pour mesurer leur empreinte environnementale et la minimiser**. Le dispositif permettra aux autorités de disposer de projets concrets pour développer, en collaboration avec leurs destinataires, des bonnes pratiques pour la conception d'applications durables.

- Etudier l'opportunité et la faisabilité de la mise en place de mécanismes incitatifs pour **favoriser l'autorégulation environnementale** des développeurs d'IA génératives (développement d'un **code de conduite**).

- Etudier l'opportunité et la faisabilité de la mise en place de mesures incitatives pour **favoriser l'approvisionnement en sources d'énergie décarbonée ou la compensation carbone** pour tous les fournisseurs de modèles d'IA génératives de taille systémique.

- Mener une étude, en s'appuyant sur le milieu académique, sur les leviers juridiques et règlementaires mobilisables pour **bâtir une régulation proportionnée**, contraignant les développeurs d'IA générative à la prise en compte de l'empreinte environnementale des systèmes qu'ils développent, **sans brider l'innovation**.

# Bibliographie